\documentclass[10pt]{article}
\usepackage{geometry}                
\usepackage{amsmath}
\usepackage{graphicx}
\usepackage{amssymb}
\usepackage{epstopdf}
\usepackage{wrapfig}

\renewcommand{\thefootnote}{\fnsymbol{footnote}}

\title{Study of fermion pair productions at the ILC with center-of-mass energy of 250 GeV}
\author{$^1$Hiroaki Yamashiro, $^1$Kiyotomo Kawagoe, $^1$Taikan Suehara, $^1$Tamaki Yoshioka, \\$^2$Keisuke Fujii, $^2$Akiya Miyamoto \\  \\Kyushu University$^1$,  KEK$^2$}
\date{}                                           

\begin{document}
\maketitle \footnote[0]{Talk presented at the International Workshop on Future Linear Colliders (LCWS2017), Strasbourg, France, 23-27 October 2017. C17-10-23.2}

\renewcommand{\thefootnote}{\arabic{footnote}}

\begin{abstract}

Precise measurements of electroweak processes at the International Linear Collider (ILC) will provide unique opportunities
to explore new physics beyond the Standard Model. Fermion pair productions are sensitive
to a new contact interaction or a new heavy gauge boson by comparing cross section and
angular distribution with expectations of the new physics models.
In this proceedings we report a simulation study of fermion pair productions
at a center-of-mass energy of 250 GeV, 
with a focus on lepton pairs, to demonstrate the potential of the first phase of the ILC.
\end{abstract}

\section{Introduction}

The
International Linear Collider (ILC) is a next generation $e^+e^-$ linear collider.
As recently proposed \cite{scenario},
the first stage of the ILC will be operated at a center-of-mass energy ($\sqrt{s}$) of 250~GeV,
mainly targeting precise measurements of the Higgs boson.
However, the ILC is also a unique facility to probe new physics not only by direct searches
of colorless new particles, but also indirectly by precise measurements of Standard Model (SM) processes.
The fermion pair production $e^+e^- \rightarrow f\bar{f}$ is one of such processes,
and any deviations in the total and differential cross sections from the expectations of the SM
can be regarded as an indirect evidence for new physics.
Unlike the case at the Large Hadron Collider (LHC), both theoretical calculations
and experimental measurements of this process can be done with O(0.1\%) precision because 
the production and decay of the $Z$ boson are purely electroweak.
We conducted a full simulation study of the $e^+e^- \rightarrow \ell^+\ell^- ~ (\ell = e, \mu ~\mathrm{and}~ \tau)$
at $\sqrt{s}=250$~GeV, and discuss possibility to discover new physics and identify physics models.



\section{Simulation Conditions}

We utilized ILCSoft\cite{ILCSoft} package of version v01-16-02-p1 for this study.
The event generation was done by WHIZARD~\cite{whi} with PYTHIA~\cite{pyt} for hadronization.
The full Monte Carlo simulation was done with Mokka  based on Geant4 \cite{G4} framework
with the reference geometry of the International Large Detector (ILD) concept used in the
studies of Detailed Baseline Design report \cite{RDR}, ILD$\_$v1$\_$05 model.
The model includes silicon pixel and strip detectors,
a time projection chamber, precisely segmented
electromagnetic and hadron calorimeters (ECAL and HCAL) and a 3.5 Tesla solenoid magnet.
Event reconstruction was done with Marlin processors \cite{Mar}, including
tracking and particle flow reconstruction by PandoraPFA algorithm~\cite{PFA} to obtain
track-cluster matching.
For the lepton tagging, simple criteria based on cluster energies in both ECAL and HCAL
and track momentum are used. Details are described in Section 3.

We assume the 250 GeV stage of H20 running scenario~\cite{scenario} with a total luminosity of 2000 fb$^{-1}$,
with 45\% $e^-_Le^+_R$ and 45\% $e^-_Re^+_L$ polarization,
where $|P(e^-)| = 0.8$ and $|P(e^+)| = 0.3$ are assumed.
We included all 2-fermion and 4-fermion pure-leptonic final states and 2-fermion hadronic
final states as background. The statistics of the Monte Carlo samples correspond to 11 to 680 fb$^{-1}$,
which is not the full statistics of H20 staging scenario, but we applied event weights to
recover the statistics.

For $e^+e^-\rightarrow e^+e^-$ (Bhabha) events, we applied a preselection of
$|\cos\theta| < 0.97$ for each track and energy sum of $e^+e^-$ more than 200 GeV
to avoid too much event simulation.

\section{Event Selection}

Here we select the $e^+e^- \rightarrow \ell^+\ell^-$ by following selection criteria,
where 
$e^+e^-$ and $\mu^+\mu^-$ final states are selected similarly except the lepton tagging,
while a different strategy is applied for selection of $\tau^+\tau^-$ final states.

The $e^+e^-$ and $\mu^+\mu^-$ events have two lepton tracks and basically nothing else in their final states.
Firstly we select a positive and a negative tracks with highest energies,
and their energies must be more than 10 GeV.
Then, we apply a lepton tagging criteria for each track based on its energy calculated from the track curvature $E_\mathrm{tr}$,
reconstructed energy deposits in ECAL ($E_\mathrm{ECAL}$) and HCAL($E_\mathrm{HCAL}$).
Electrons can be identified by that most of their energy is deposited at the ECAL than HCAL,
and muons can be identified by penetration of calorimeters.
For electrons, $(E_\mathrm{ECAL}+E_{\mathrm{HCAL}})/E_{\mathrm{tr}} > 0.6$ and $E_{\mathrm{ECAL}}/(E_{\mathrm{ECAL}}+E_{\mathrm{HCAL}}) > 0.9$
are required, while for muons, $(E_\mathrm{ECAL}+E_\mathrm{HCAL})/E_\mathrm{tr} < 0.6$ and $E_\mathrm{ECAL}/(E_\mathrm{ECAL}+E_\mathrm{HCAL}) < 0.5$ are required. Both of the tracks should pass the criteria to be recognized 
as $e^+e^-$ and $\mu^+\mu^-$ final states.
After that, we apply a kinematic cut that energy sum of the two tracks $E_{\mathrm{sum}}$ must be larger than 230 GeV 
and both of the tracks should
be in an angular region $|\cos\theta| < 0.95$, where $\theta$ is the angle with respect to the beam axis.
The first selection rejects most of $W^+W^-$, $\tau^+\tau^-$ and $Z\gamma\rightarrow\ell^+\ell^-\gamma$ events and the second selection rejects mostly $t$-channel $e^+e^-$ events.
The cut statistics are shown in Tables \ref{tab:Cut_e} and~\ref{tab:Cutmu}.

\begin{table}[h]
\begin{center}
  \begin{tabular}{|l||r|r|r||r|r|r|} \hline
Selection of $e^+e^-\rightarrow e^+e^-$&\multicolumn{3}{|l|}{$ e^-_L e^+_R $}&\multicolumn{3}{|l|}{$ e^-_R e^+_L $} \\ \hline
&signal&$2f $~bkg.&$4f $~bkg.&signal&$2f $~bkg.&$4f $~bkg.\\ \hline
All events after preselection&216M&11.7M &44.9M &210M &9.34M&1.24M\\ \hline
Two tracks with $E_\mathrm{tr} >$ 10 GeV &212M&8.75M&3.65M&207M&6.77M&966k\\ \hline
Electron tagging &209M&410k&849k&204M&309k&565k\\ \hline
$E_\mathrm{sum} ~> 230 $ ~ GeV &105M&44&40.0k &102M &29 &38.7k \\ \hline
$| \cos \theta| < 0.95$  &55.8M&33&9.95k&54.5M&12&9.56k\\ \hline
  \end{tabular}
\end{center}
\caption{Cut statistics of $e^+e^-\rightarrow e^+e^-$ channel.
The ``Two tracks with $E_\mathrm{tr} > 10$ GeV" row includes the selection
of having two oppositely-charged particles. Details of electron tagging is described in the text.
}
\label{tab:Cut_e}
\end{table}

\begin{table}[h]
\begin{center}
  \begin{tabular}{|l||r|r|r||r|r|r|} \hline
Selection of $e^+e^-\rightarrow \mu^+\mu^-$&\multicolumn{3}{|l|}{$ e^-_L e^+_R $}&\multicolumn{3}{|l|}{$ e^-_R e^+_L $} \\ \hline
&signal&$2f $~bkg.&$4f $~bkg.&signal&$2f $~bkg.&$4f $~bkg.\\ \hline
All events after preselection&6.13M&221M &44.9M&5.01M&214M &1.24M \\ \hline
Two tracks with $E_\mathrm{tr} >$ 10 GeV &5.03M&216M&3.65M&3.94M&208M&966k\\ \hline
Muon tagging&3.96M&79.6k &550k &3.11M&57.9k&70.3k\\ \hline
$E_\mathrm{sum} ~> 230 $ ~GeV &1.61M&11 &743 &1.34M&1&596\\ \hline
$| \cos \theta| < 0.95$  &1.52M&11&71&1.27M&1&578\\ \hline
  \end{tabular}
\end{center}
\caption{Cut statistics of $e^+e^-\rightarrow \mu^+ \mu^-$ channel.
The ``Two tracks with $E_\mathrm{tr} > 10$ GeV" row includes the selection
of having two oppositely-charged particles. Details of muon tagging is described in the text.
}
\label{tab:Cutmu}
\end{table}

The $\tau$ pair final state needs slightly different selection since a $\tau$ lepton decays to multiple particles.  So we need ``tau clustering".
Our clustering method is described in \cite{TC}, which is basically clustering of particles with an invariant mass less than 2 GeV. We select a positive and a negative clusters with highest energies as the $\tau$ candidates. We require the cluster energy
$E_\mathrm{clu} >10 $~GeV for each of them.
Then the opening angle between the clusters ($\theta_{cc}$) is required to be larger than 178 degree to reject $Z\gamma$ events, the visible energy should be between 50 and 200 GeV to separate $e^+e^-$ and $\mu^+ \mu^-$ events, and
the $\tau$ clusters are in the angular region
$|\cos\theta| < 0.95$ to reject $t$-channel events. The cut statistics are shown in Table \ref{tab:CutTau}.

\begin{table}[h]
\begin{center}
  \begin{tabular}{|l||r|r|r||r|r|r|} \hline

Selection of $e^+e^-\rightarrow \tau^+\tau^-$&\multicolumn{3}{|l|}{$ e^-_L e^+_R $}&\multicolumn{3}{|l|}{$e^-_R e^+_L $} \\ \hline
&signal&$2f $~bkg.&$4f $~bkg.&Signal&$2f $~bkg.&$4f $~bkg.\\ \hline
All events&5.56M &222M &44.9M&4.33M&215M &1.24M \\ \hline
Two tau clusters with $E_\mathrm{clu} >$ 10 GeV &1.89M&132M&2.02M &1.44M &128M&238k\\ \hline
$\theta_{cc}$ $>$ 178 deg. &705k&11.6k &46.6k&568k&112M &40.7k \\ \hline
$50 ~ \mathrm{GeV}<E_{\mathrm{vis}} < 200 ~\mathrm{GeV} $&658k&61.2M &8.30k &530k&59.4M&4.10k\\ \hline
$| \cos \theta| < 0.95$  &587k&12.9k&3.55k&455k&13.6k &304\\ \hline
  \end{tabular}
\end{center}
\caption{Cut statistics of $e^+e^-\rightarrow \tau^+ \tau^-$ channel.
The ``Two tau clusters with $E_\mathrm{tr} > 10$ GeV" row includes the selection
of having two oppositely-charged clusters. 
}
\label{tab:CutTau}
\end{table}

\section{Analysis}

\begin{figure}[h]
	\centering
	\begin{minipage}{0.45\columnwidth}
		\centering
		\includegraphics[width=70mm,height=80mm,keepaspectratio,clip,angle=0]{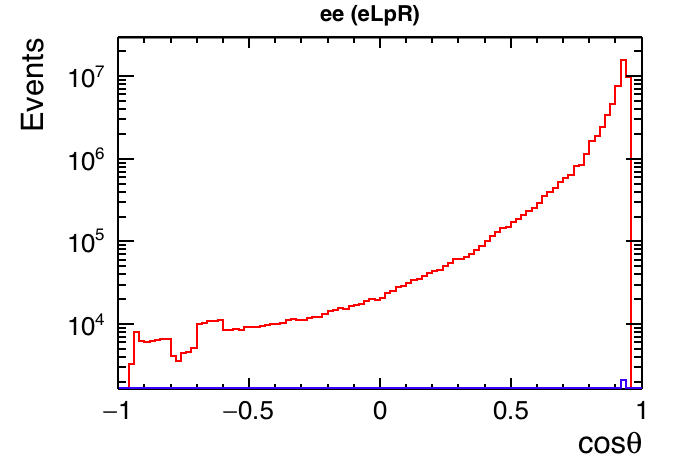}
	\end{minipage}
	\centering
	\begin{minipage}{0.45\columnwidth}
		\centering
		\includegraphics[width=70mm,height=80mm,keepaspectratio,clip,angle=0]{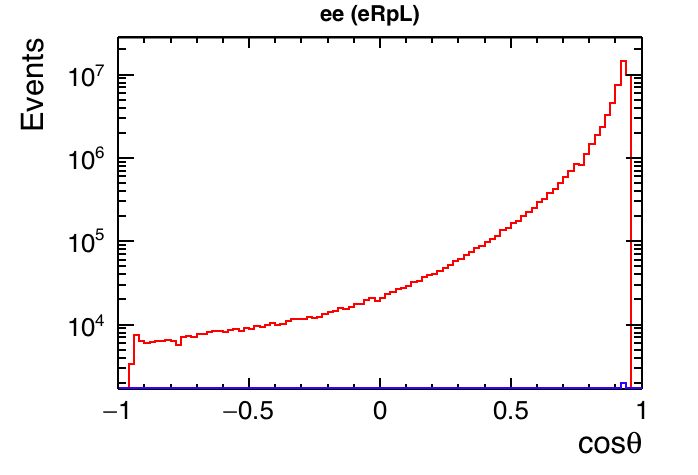}
	\end{minipage}
	
	\centering
	\begin{minipage}{0.45\columnwidth}
		\centering
		\includegraphics[width=70mm,height=80mm,keepaspectratio,clip,angle=0]{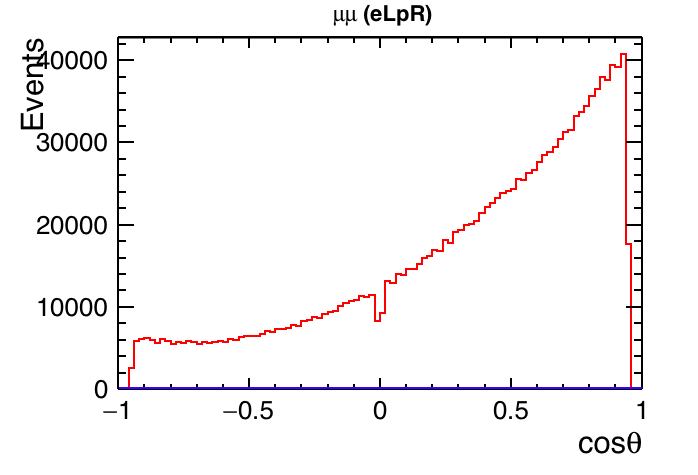}
	\end{minipage}	
		\centering
	\begin{minipage}{0.45\columnwidth}
		\centering
		\includegraphics[width=70mm,height=80mm,keepaspectratio,clip,angle=0]{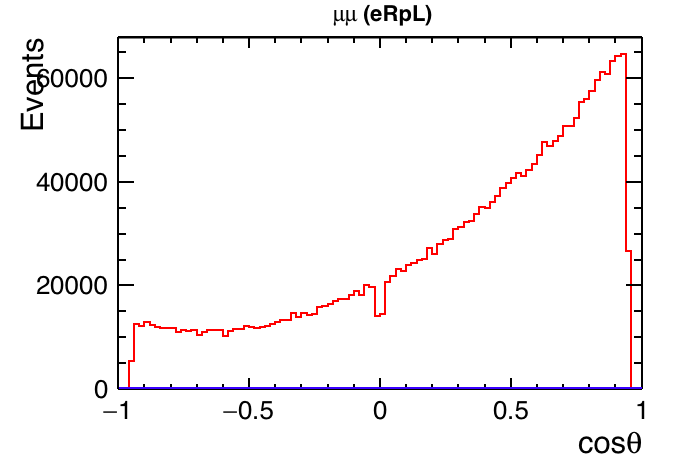}
	\end{minipage}
	
		\centering	
	\begin{minipage}{0.45\columnwidth}
		\centering
		\includegraphics[width=70mm,height=80mm,keepaspectratio,clip,angle=0]{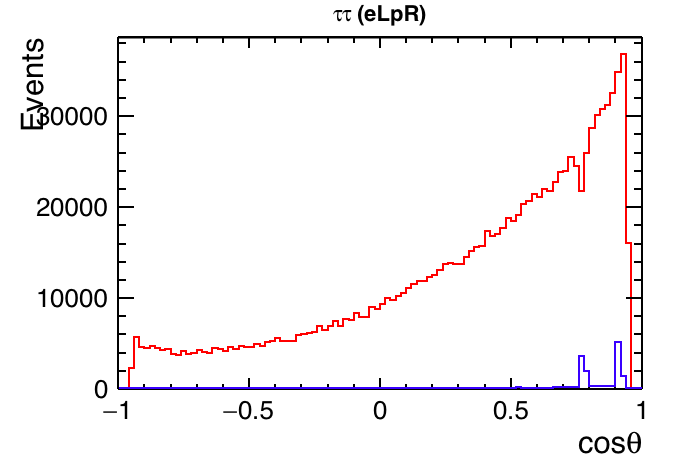}
	\end{minipage}	
		\centering
	\begin{minipage}{0.45\columnwidth}
		\centering
		\includegraphics[width=70mm,height=80mm,keepaspectratio,clip,angle=0]{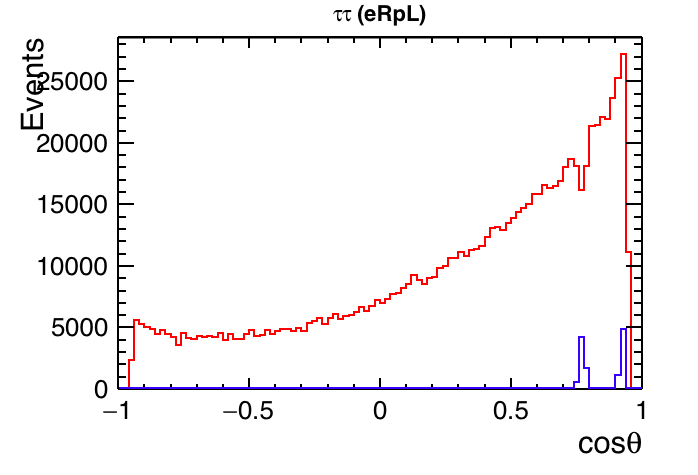}
	\end{minipage}
		\caption{ Angular distributions after the event selections.
Left (right) figures show the distributions of $e^-_Le^+_R$ ($e^-_Re^+_L$) polarization.
The upper, the middle and the lower figures show $e^+e^-$, $\mu^+\mu^-$ and $\tau^+\tau^-$ channels, respectively.
The $e^+e^-$ channels are shown in logarithmic scales while the other channels are shown in linear scales.
Red (blue) lines show distributions of signal (background) events. }
		\label{fig:Ang}	
\end{figure}

Figure \ref{fig:Ang} shows the angular distributions after the event selection.
These distributions are used to evaluate the precision of the ILC measurements on $i$-th $\cos\theta$ bin, $\delta\sigma_i / \sigma_i(\mathrm{SM})$, by
\begin{equation}
	\frac{\delta\sigma_i}{\sigma_i(\mathrm{SM})}  = \frac{\sqrt{S_i + N_i}}{S_i}
\end{equation}
where $S_i$ and $N_i$ are the number of signal and background events in each bin.  In this study systematic uncertainty is not considered.
 
Here we investigated possibility to search for two types of new physics models based on the obtained precision.
The first is $Z'$ models \cite{Zp}, where $Z'$ is an additional neutral vector gauge boson coupled to SM fermions.
The coupling constants differ depending on models, and we used SSM (Sequential Standard Model), 
and 
$E_6$ models. The SSM assumes the same coupling constants as SM $Z$.
On the other hand, the $E_6$ is a string-motivated model which naturally introduces $Z'$ as a linear combination of the two extra $U(1)$ gauge bosons $Z_{\psi}$ and $Z_{\chi}$ : $Z' =Z_{\chi}\cos\beta + Z_{\psi}\sin\beta$. We investigated three $\beta$ parameters: $\beta = 0$ ($\chi$ model), $\beta = \pi /2 $ ($\psi$ model) and $ \beta =\pi - \mathrm{arctan} \sqrt{5/3}$ ($\eta$ model).
ALR (Alternative Left-Right symmetric) is another model also introduced from $E_6$, which gives extra $SU(2)_R$ in addition to SM $SU(2)_L$. This introduces an additional $Z_R$ boson phenomenologically treated as $Z'$, which behaves like SM $Z$, but gives different couplings to SM particles.

The LHC experiments already gave lower limits at around 4.5 TeV ($Z'\rightarrow e^+e^-$ and $\mu^+\mu^-$ combined) 
and around 2.4 TeV ($Z'\rightarrow\tau^+\tau^-$) \cite{ATLAS, CMS} with direct reconsturction of the mass peak assuming the SSM model,
but the LHC experiments have not given limits assuming other models.

The new physics model is a generic search of a WIMP (weakly-interacting massive particle) dark matter \cite{WIMP}.
In the $e^+e^- \rightarrow f\bar{f}$ process,  WIMP ($\chi$) can be introduced in a $Z\rightarrow\chi\chi\rightarrow Z$ loop diagram,
which gives a correction to the coupling constant. The correction basically only depends on a group structure,
spin and mass of $\chi$ and is independent of model details. We investigated three well-motivated types of WIMPs:
wino ($SU(2)_L$ triplet and $U(1)_Y$ hypercharge of 0), Higgsino ($SU(2)_L$ doublet and $U(1)_Y$ hypercharge of $\pm$1/2) and Minimal Dark Matter ($SU(2)_L$ pentet and $U(1)_Y$ hypercharge of 0).

To investigate the performance of search for WIMP in these models, 
we obtained the deviation of $e^+e^- \rightarrow f\bar{f}$ cross section: 
$\delta\sigma_i (\mathrm{BSM})/\sigma_i(\mathrm{SM})$ by theoretical prediction and obtain the $\chi^2$ value for each model as
\begin{equation}
\chi^2(\mathrm{BSM}) = \sum_i\left\{\left(\frac{\delta\sigma_i(\mathrm{BSM})}{\sigma_i(\mathrm{SM})}
/ \frac{S_i}{\sqrt{S_i + N_i}}\right)^2 + 1\right\},
\end{equation}
to calculate the rejection probability of each model with pure-SM distribution.

\begin{figure}[h]
	\centering
	\begin{minipage}{0.45\columnwidth}
		\centering
		\includegraphics[width=60mm,height=70mm,keepaspectratio,clip,angle=0]{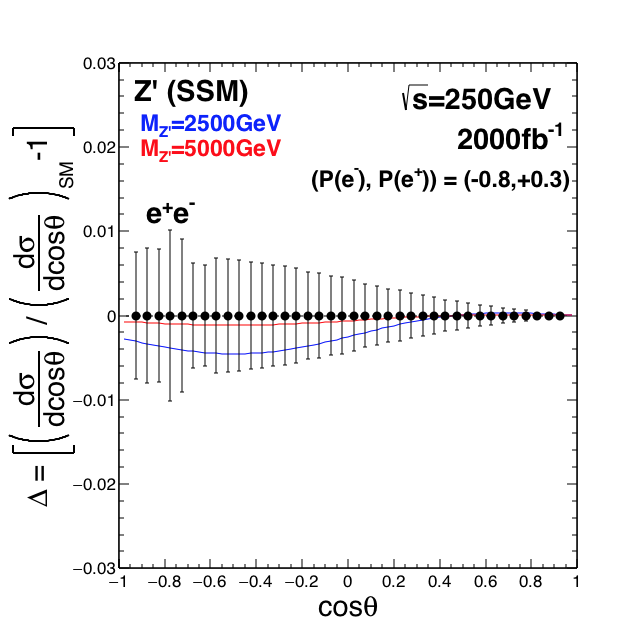}
	\end{minipage}
	\centering
	\begin{minipage}{0.45\columnwidth}
		\centering
		\includegraphics[width=60mm,height=70mm,keepaspectratio,clip,angle=0]{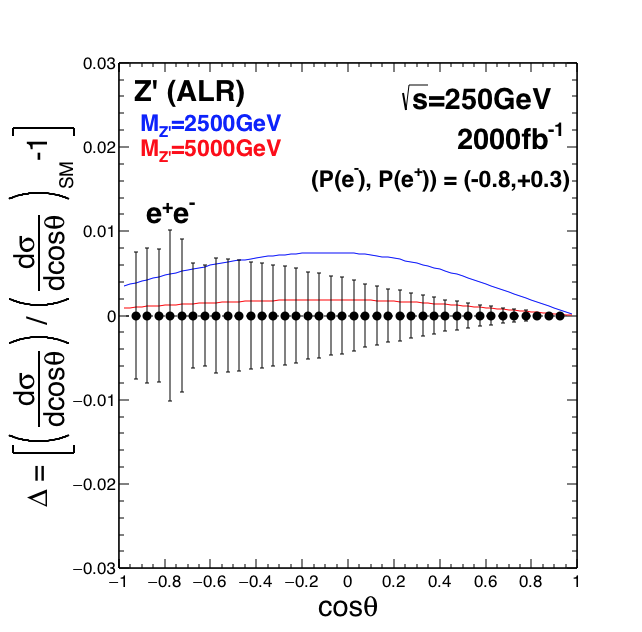}
	\end{minipage}

	\centering
	\begin{minipage}{0.45\columnwidth}
		\centering
		\includegraphics[width=60mm,height=70mm,keepaspectratio,clip,angle=0]{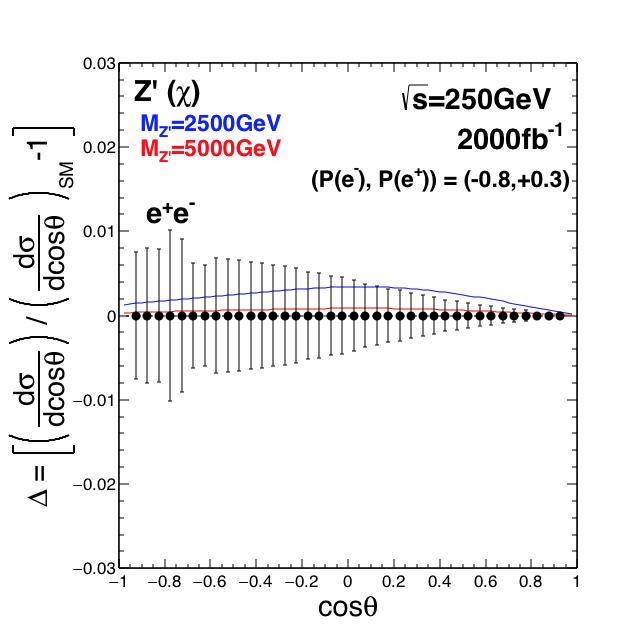}
	\end{minipage}
	\centering
	\begin{minipage}{0.45\columnwidth}
		\centering
		\includegraphics[width=60mm,height=70mm,keepaspectratio,clip,angle=0]{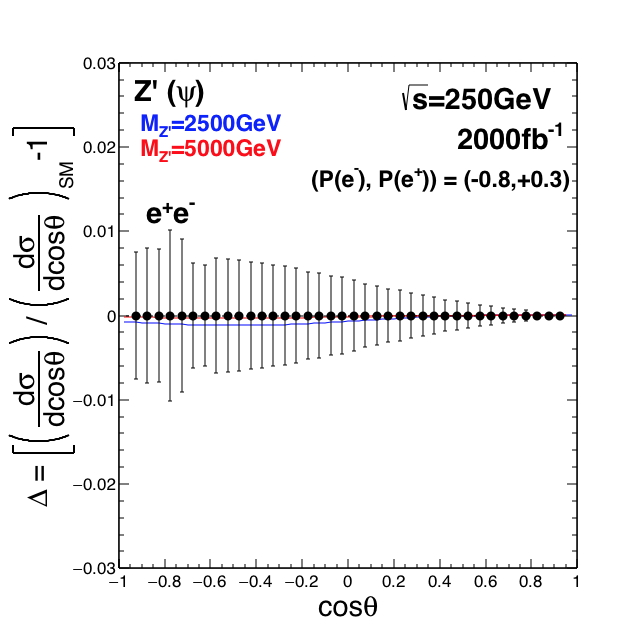}
	\end{minipage}
	\centering
	\includegraphics[width=60mm,height=70mm,keepaspectratio,clip,angle=0]{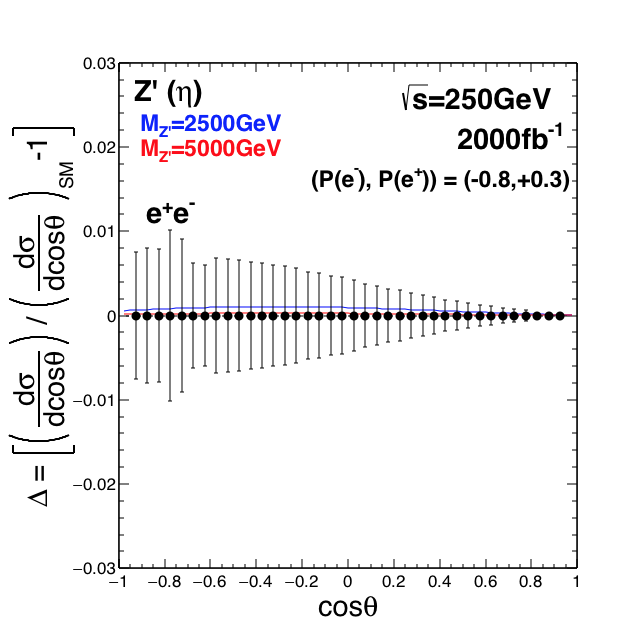}
	
	\caption{Effect of $Z'$ of SSM and $E_6$ models
as deviation of the differential cross section of $e^+e^-\rightarrow e^+e^-$ from the SM.
Red (blue) lines show the deviation by $Z'$ mass of 2.5 (5.0) TeV.
The expected precision of measurements at $\sqrt{s}=250$~GeV is shown in the error bars,
assuming H20 staging scenario written in the text.}
	\label{fig:dist1}
\end{figure}

\begin{figure}[h]
	\centering
	\begin{minipage}{0.45\columnwidth}
		\centering
		\includegraphics[width=60mm,height=70mm,keepaspectratio,clip,angle=0]{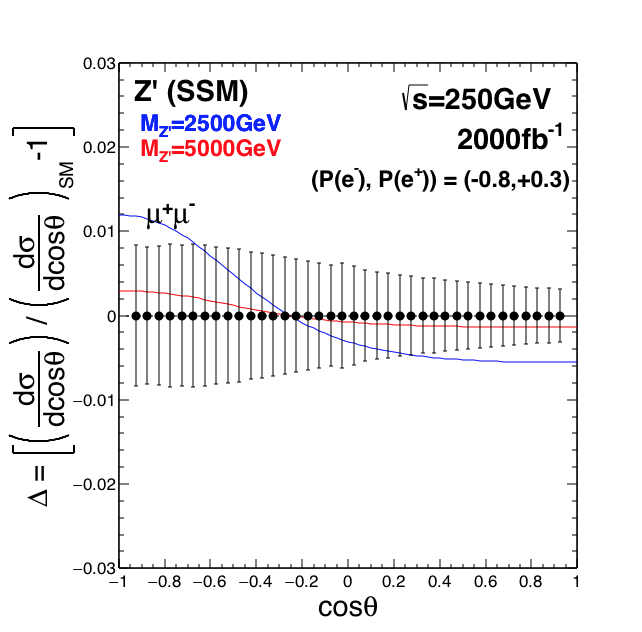}
	\end{minipage}
	\centering
	\begin{minipage}{0.45\columnwidth}
		\centering
		\includegraphics[width=60mm,height=70mm,keepaspectratio,clip,angle=0]{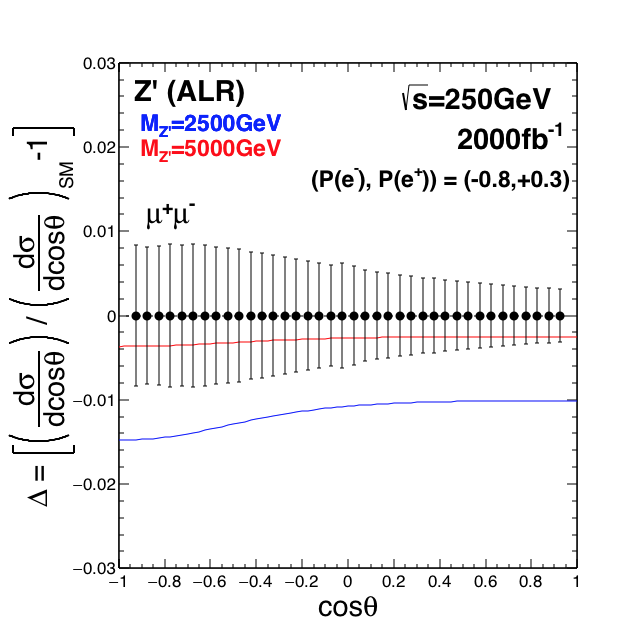}
	\end{minipage}

	\centering
	\begin{minipage}{0.45\columnwidth}
		\centering
		\includegraphics[width=60mm,height=70mm,keepaspectratio,clip,angle=0]{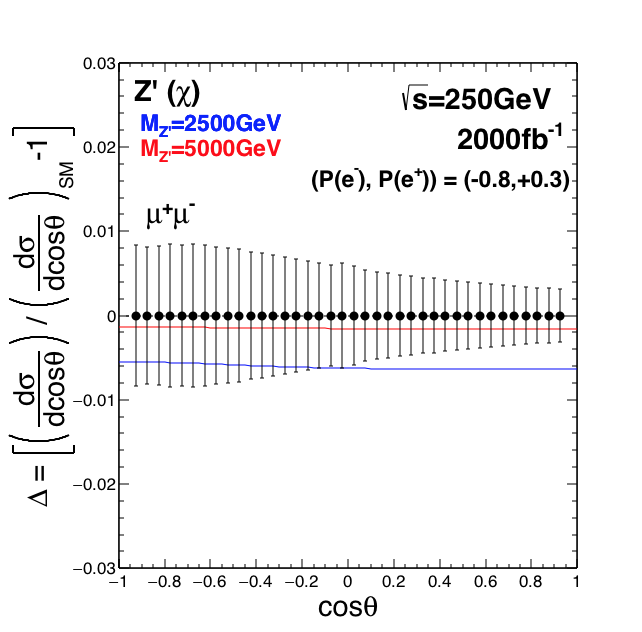}
	\end{minipage}
	\centering
	\begin{minipage}{0.45\columnwidth}
		\centering
		\includegraphics[width=60mm,height=70mm,keepaspectratio,clip,angle=0]{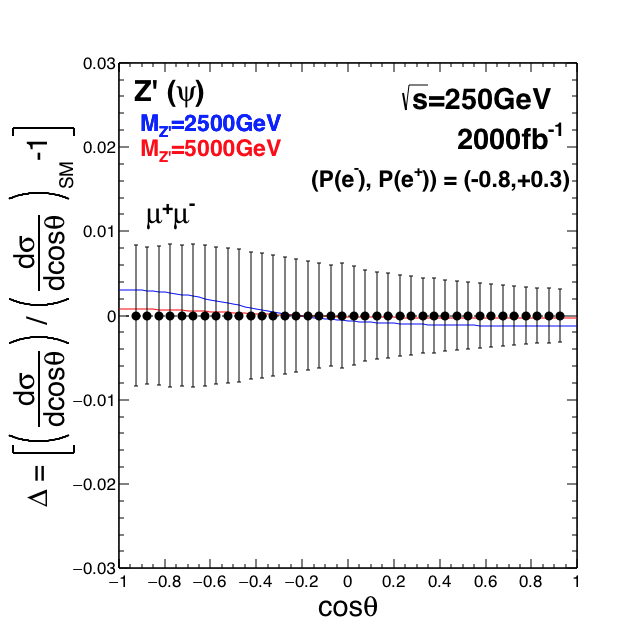}
	\end{minipage}
	\centering
	\includegraphics[width=60mm,height=70mm,keepaspectratio,clip,angle=0]{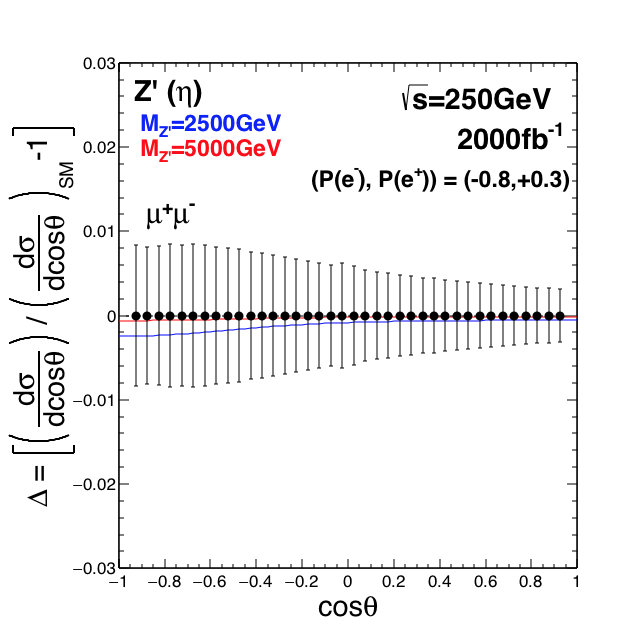}
	
	\caption{Effect of $Z'$ of SSM and $E_6$ models
as deviation of the differential cross section of $e^+e^-\rightarrow \mu^+\mu^-$ from the SM.
Red (blue) lines show the deviation by $Z'$ mass of 2.5 (5.0) TeV.
The expected precision of measurements at $\sqrt{s}=250$~GeV is shown in the error bars,
assuming H20 staging scenario written in the text.}
	\label{fig:dist2}
\end{figure}

\begin{figure}[h]
	\centering
	\begin{minipage}{0.45\columnwidth}
		\centering
		\includegraphics[width=60mm,height=70mm,keepaspectratio,clip,angle=0]{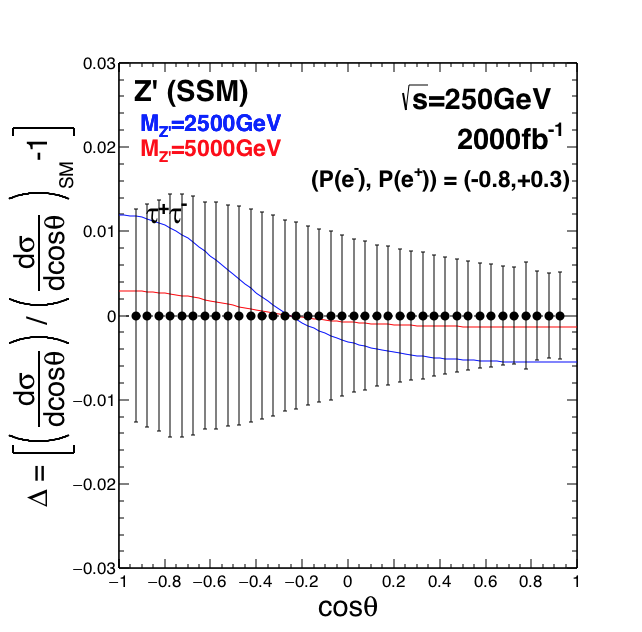}
	\end{minipage}
	\centering
	\begin{minipage}{0.45\columnwidth}
		\centering
		\includegraphics[width=60mm,height=70mm,keepaspectratio,clip,angle=0]{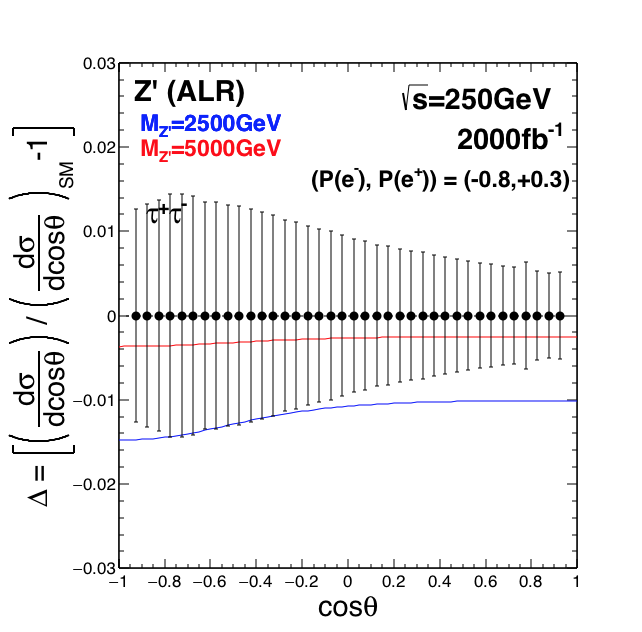}
	\end{minipage}

	\centering
	\begin{minipage}{0.45\columnwidth}
		\centering
		\includegraphics[width=60mm,height=70mm,keepaspectratio,clip,angle=0]{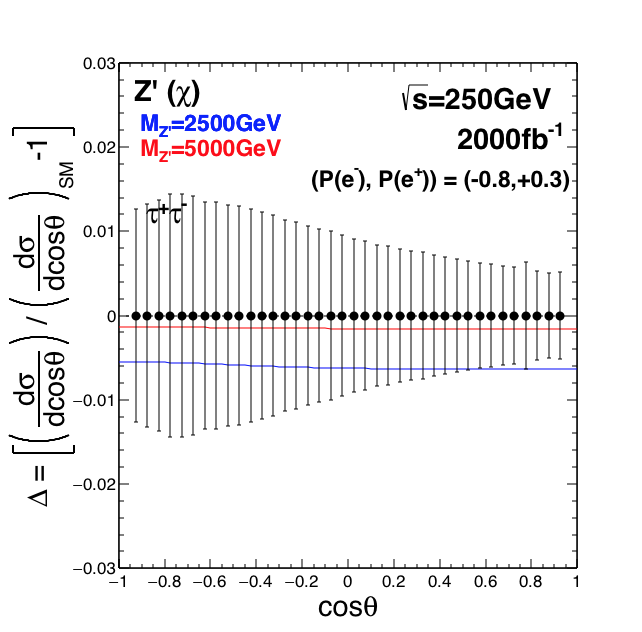}
	\end{minipage}
	\centering
	\begin{minipage}{0.45\columnwidth}
		\centering
		\includegraphics[width=60mm,height=70mm,keepaspectratio,clip,angle=0]{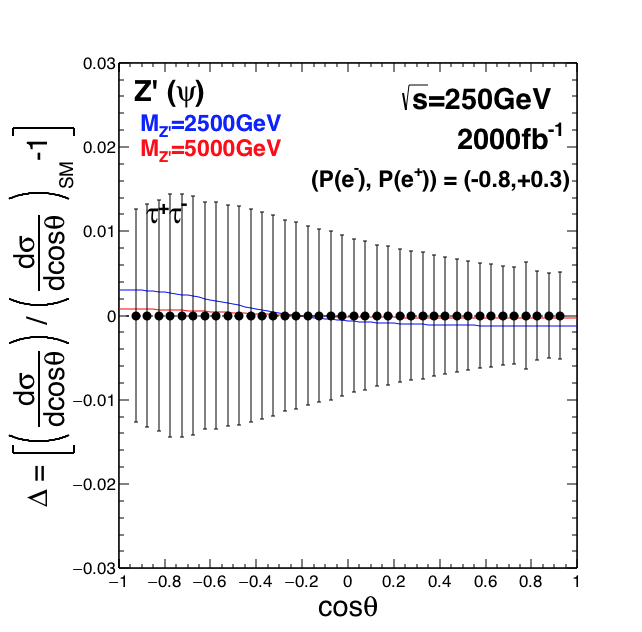}
	\end{minipage}
	\centering
		\includegraphics[width=60mm,height=70mm,keepaspectratio,clip,angle=0]{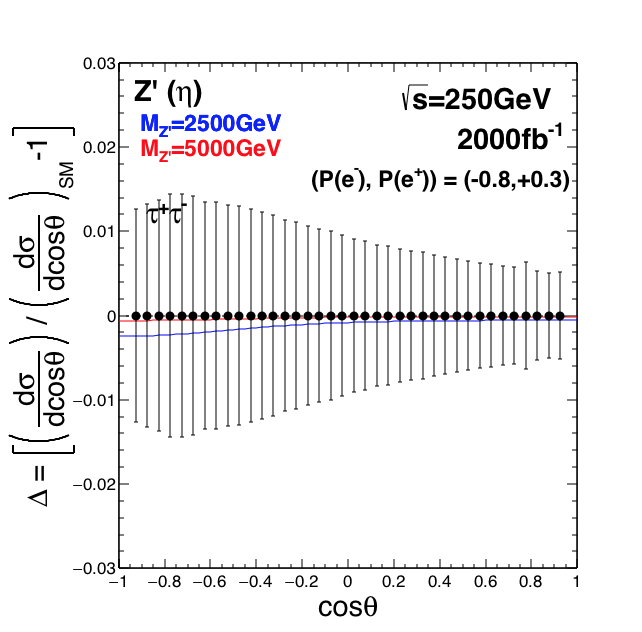}
	
	\caption{Effect of $Z'$ of SSM and $E_6$ models
as deviation of the differential cross section of $e^+e^-\rightarrow \tau^+\tau^-$ from the SM.
Red (blue) lines show the deviation by $Z'$ mass of 2.5 (5.0) TeV.
The expected precision of measurements at $\sqrt{s}=250$~GeV is shown in the error bars,
assuming H20 staging scenario written in the text.}
	\label{fig:dist3}
\end{figure}

Figures \ref{fig:dist1}, \ref{fig:dist2} and \ref{fig:dist3} shows the $\delta\sigma_i (\mathrm{BSM})/\sigma_i(\mathrm{SM})$ and
$\delta\sigma_i (\mathrm{BSM})/\sigma_i(\mathrm{SM})$ of $e^+e^- \rightarrow e^+e^-$, $\mu^+\mu^-$ and $\tau^+\tau^-$ final states.
Here we only show the distribution on $e^-_Le^+_R$ which is much more powerful than $e^-_Re^+_L$ polarization.
The plots show that 2.5 TeV $Z'$ is easily discovered for most of the models while 5.0 TeV $Z'$ is hard to identify.

\begin{figure}[h]
	\centering
	\begin{minipage}{0.6\columnwidth}
		\centering
		\includegraphics[width=100mm,height=120mm,keepaspectratio,clip,angle=0]{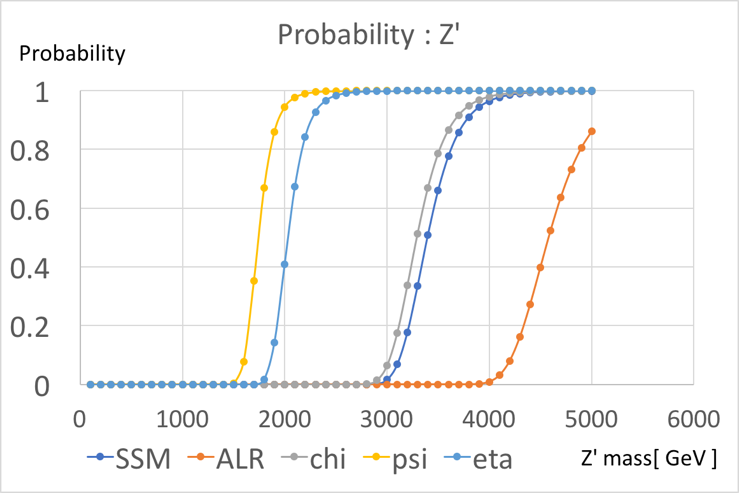}
	\end{minipage}
	\centering
	\begin{minipage}{0.6\columnwidth}
		\centering
		\includegraphics[width=100mm,height=120mm,keepaspectratio,clip,angle=0]{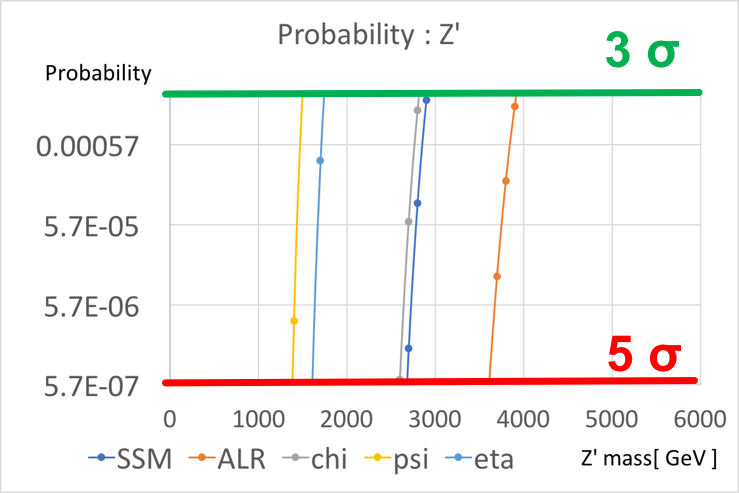}
	\end{minipage}
		\caption{Probability of the distributions consistent to SM under the deviation by $Z'$ models,
calculated by the $\chi^2$ defined in the text. The left figure shows the dependence of
probability on the $Z'$ mass on each model, and the right figure shows the minimum $Z'$ mass
which can be detected as 3$\sigma$ and 5$\sigma$ deviation from the SM.
 $e$, $\mu$ and $\tau$ channels are combined.}
		\label{fig:probZ}
\end{figure}

The calculated probability based on the $\chi^2$ with various $Z'$ masses is shown in Figure \ref{fig:probZ}.
The obtained $3 \sigma$ limit of the model rejection is shown in Table \ref{tab:MR}.

\begin{table}[h]
\begin{center}
  \begin{tabular}{|c|c|} \hline
      $Z^{\prime}$ model& mass reach at $3\sigma$\\ \hline
SSM & $  2.8 $TeV \\ \hline 
ALR & $ 4.0 $TeV \\ \hline 
$\chi$ & $  2.9 $TeV \\ \hline 
 $\psi $ &$  1.4 $TeV \\ \hline 
   $\eta$ & $  1.8 $TeV \\ \hline 
  
  \end{tabular}
\end{center}
\caption{The minimal $Z'$ mass observed as the 3$\sigma$ deviation from the SM by $e^+e^-\rightarrow \ell^+\ell^-$ measurements of 250  GeV ILC.}
\label{tab:MR}
\end{table}

\begin{figure}[h]
 \centering
   \includegraphics[width=160mm,keepaspectratio]{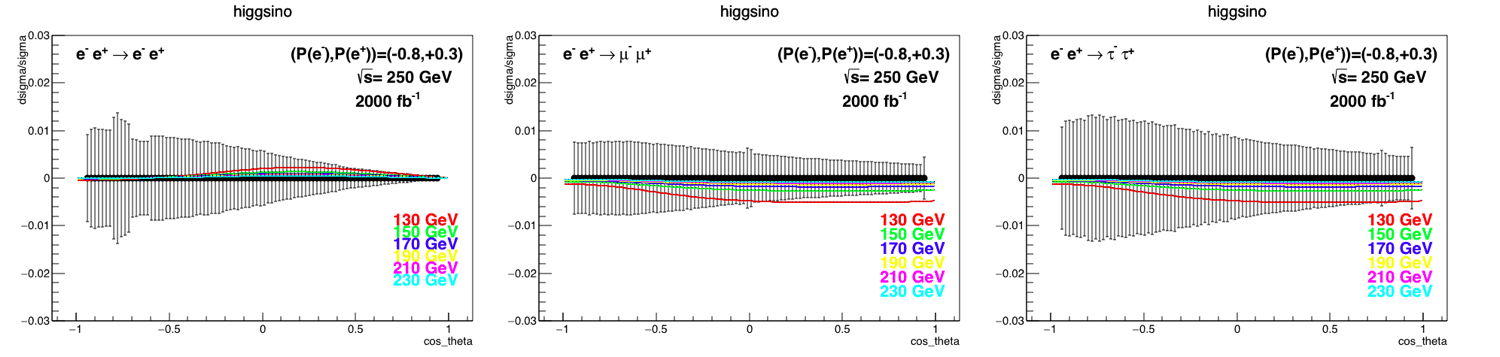}

\centering
  \includegraphics[width=160mm,keepaspectratio]{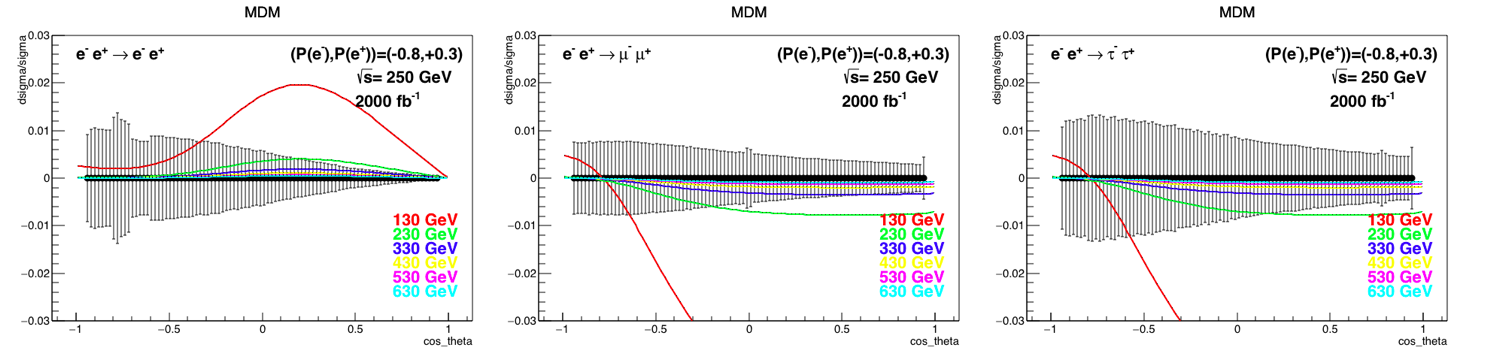}
  \centering
    \includegraphics[width=160mm,keepaspectratio]{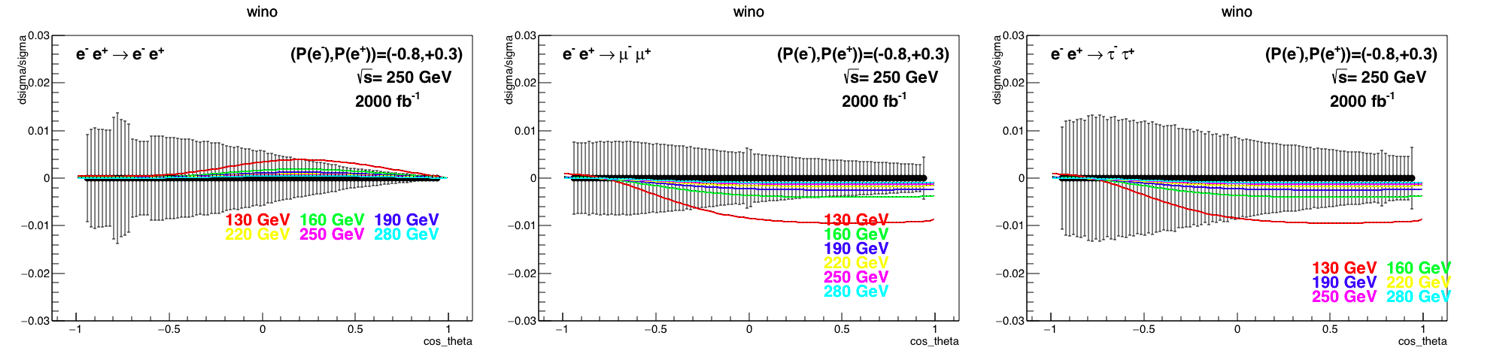}
 \caption{Effect of WIMP models as deviation of the differential cross section of $e^+e^-\rightarrow \ell^+\ell^-$ from the SM.
The upper, the middle and the lower figures show the Higgsino ($n = 2, Y = \pm1/2$), wino ($n = 3, Y = 0$) and
Mimimal Dark Matter ($n = 5, Y = 0$) where $n$ is $SU(2)_L$ $n$-plet and $Y$ is the $U(1)_Y$ hypercharge, respectively.
The left, the middle and the right figures show deviation of $e^+e^-$, $\mu^+\mu^-$ and $\tau^+\tau^-$ channels, respectively.
The error bars show the expected precision at $\sqrt{s}=250$~GeV. }
 \label{fig:MDM}
\end{figure}

\begin{figure}[h]
	\centering
	\begin{minipage}{0.6\columnwidth}
		\centering
	\includegraphics[width=100mm,height=120mm,keepaspectratio,clip,angle=0]{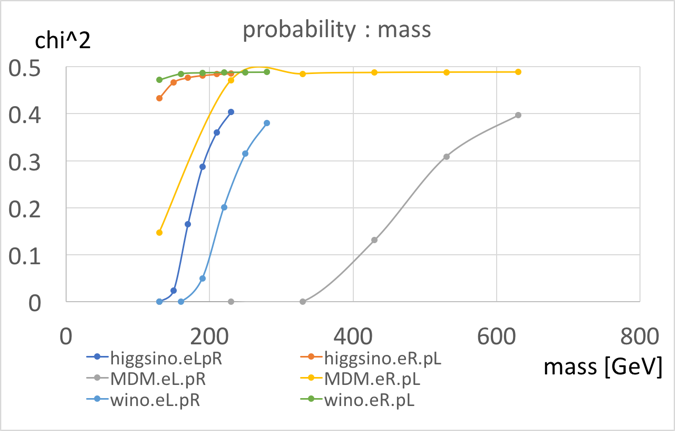}
	\end{minipage}
	\centering
	\begin{minipage}{0.6\columnwidth}
		\centering
		\includegraphics[width=100mm,height=120mm,keepaspectratio,clip,angle=0]{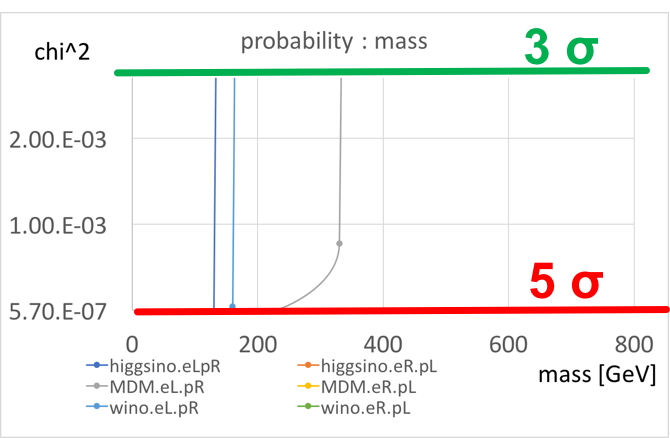}
	\end{minipage}
	\caption{ Probability of the distributions consistent to SM under the deviation by WIMP models,
calculated by the $\chi^2$ defined in the text. The left figure shows the dependence of
probability on the WIMP mass on each model, and the right figure shows the minimum WIMP mass
which can be detected as 3$\sigma$ and 5$\sigma$ deviation from the SM.
 $e$, $\mu$ and $\tau$ channels are combined.}
	\label{fig:probW}
\end{figure}

Figure \ref{fig:MDM} is similar to Figures \ref{fig:dist1}, \ref{fig:dist2} and \ref{fig:dist3}, but  for
generic WIMP search of various WIMP masses. The probability distribution and $3 \sigma$ limit is shown
in Figure \ref{fig:probW} and Table \ref{tab:kkkk}. It shows that the mass reach is higher than the beam energy, which means
we can extend the searching power of WIMP by using this measurement.

\begin{table}[h]
\begin{center}
  \begin{tabular}{|c|c|} \hline
    WIMP Model  & mass reach at $3\sigma$\\ \hline
 higgsino $e^-_L e^+_R $ & $  150 $GeV \\ \hline 
 MDM $e^-_L e^+_R $ & $  330 $GeV \\ \hline 
  wino $e^-_L e^+_R $ & $  150 $GeV \\ \hline 
  
  \end{tabular}
\end{center}
\caption{The minimal WIMP mass observed as the 3$\sigma$ deviation from the SM by $e^+e^-\rightarrow \ell^+\ell^-$ measurements
at $\sqrt{s}=250 $~GeV at the ILC.}
\label{tab:kkkk}
\end{table}


\section{Summary and Prospects}
We investigated the $e^+e^-\rightarrow \ell^+\ell^-$ final states at $\sqrt{s}=250$~GeV
for a new physics study. Precise measurements of total and differential cross sections
of the process can be a good probe to new physics including a $Z'$ or a WIMP.
In our study, $Z'$ can be probed up to 1.4-4.0 TeV depending on the models,
and WIMP can be searched for up to 150-330 GeV, which significantly extends the possibility
to discover these new particles from direct searches.
We are investigating more new physics models, such as
the Gauge Higgs Unification model described at \cite{GHU}
We plan to include the hadronic final states in the near future to conclude this study.
For the hadronic final states, charge identification of jets is essential,
which is a good challenge for the ILC detectors and the reconstruction software.
The performance of the jet charge identification has been studied in \cite{pjet}, which can  be
applicable to this study.

\section*{Acknowledgements}
We appreciate S.~Shirai for theoretical calculations for WIMP models, and ILD physics and software group
for the production of event samples and the support of the software and the computing environment.
This work was supported by JSPS KAKENHI Grant Number 16H02176.

\end{document}